\documentclass[aps,prl,amsmath,amsfonts,amssymb,notitlepage,twocolumn,superscriptaddress,footinbib,floatfix]{revtex4-2}

\usepackage{CJK}
\usepackage{bm}
\usepackage{soul}
\usepackage[caption=false,subrefformat=parens,labelformat=parens]{subfig}
\usepackage{color} 
\usepackage[table,dvipsnames]{xcolor}
\usepackage[colorlinks,linktocpage,bookmarks=false]{hyperref}
\usepackage{enumitem}
\usepackage{ulem}
\usepackage{xspace}

\newcommand\myshade{85}
\definecolor{myrulecolor}{RGB}{150,20,0}
\colorlet{mylinkcolor}{violet}
\colorlet{mycitecolor}{YellowOrange}
\colorlet{myurlcolor}{Aquamarine}
\hypersetup{
	linkcolor = myurlcolor!\myshade!black,
	citecolor = mycitecolor!\myshade!black,
	urlcolor = myrulecolor!\myshade!black,
} 

\usepackage{graphicx}
\graphicspath{{FIG/}}
\usepackage{multirow}
\usepackage{braket} 
\usepackage{booktabs} 

\AtBeginDocument{
	\heavyrulewidth=.08em
	\lightrulewidth=.05em
	\cmidrulewidth=.03em
	\belowrulesep=.65ex
	\belowbottomsep=0pt
	\aboverulesep=.4ex
	\abovetopsep=0pt
	\cmidrulesep=\doublerulesep
	\cmidrulekern=.5em
	\defaultaddspace=.5em
}

\newcommand{\beq}{\begin{equation}}
\newcommand{\eeq}{\end{equation}}
\newcommand{\bea}{\begin{eqnarray}}
\newcommand{\eea}{\end{eqnarray}}

\newcommand{\ft}{fragile topological\xspace}

\DeclareMathAlphabet\mathbfcal{OMS}{cmsy}{b}{n}

\newcommand\CP{\mathbb{C}P}
\newcommand\RP{\mathbb{R}P}

\renewcommand\[{\begin{equation}}
\renewcommand\]{\end{equation}}

\makeindex

\begin{document} 
\begin{CJK*}{UTF8}{gbsn} 
\title{Classification of Classical Spin Liquids: Typology and Resulting Landscape
}
\author{Han Yan (闫寒)}
\affiliation{Department of Physics and Astronomy, Rice University, Houston, TX 77005, USA}
\affiliation{Smalley-Curl Institute, Rice University, Houston, TX 77005, USA}
\author{Owen Benton} 
\affiliation{Max Planck Institute for Physics of Complex Systems, N\"{o}thnitzer Str. 38, Dresden 01187, Germany}
\author{Roderich Moessner} 
\affiliation{Max Planck Institute for Physics of Complex Systems, N\"{o}thnitzer Str. 38, Dresden 01187, Germany}
 \author{Andriy H. Nevidomskyy}
\affiliation{Department of Physics and Astronomy, Rice University, Houston, TX 77005, USA}
\date{\today}
\begin{abstract}
{
Classical spin liquids (CSL) lack long-range magnetic order and are characterized by an {extensive} ground state  degeneracy. 
We propose a classification scheme of CSLs based on the structure of the flat bands of their Hamiltonians. Depending on  absence or presence of the gap from the flat band, the CSL are classified as algebraic or fragile topological, respectively. Each category is further classified: the algebraic case by the nature of the  emergent Gauss's law at the gap-closing point(s), and the \ft case by the homotopy of the eigenvector winding around the Brillouin zone. Previously identified instances of CSLs fit snugly into our scheme, which finds a landscape where algebraic CSLs are located at transitions between \ft ones. It also allows us to present a new, simple family of models illustrating that landscape, which}  hosts both \ft and algebraic CSLs, as well as transitions between them.
\end{abstract}
\maketitle
\end{CJK*}

\noindent\textbf{\textit{Introduction.  }}
Research into magnets without long-range order has a long history, from the study of the effects of disorder in spin glasses \cite{Edwards1975,SK_glass}, to the proposal of resonating valence bond states \cite{ANDERSON1973RVB,Anderson1987} that underpins much of modern research in strongly frustrated magnets. Classical spin liquids (CSLs) can arise when the ground state manifold of a continuous spin model is {extensively} degenerate, representing the extreme limit of the consequences of frustration, when fluctuations between ground states preclude any form of order altogether \cite{Villain1979,MC_pyro_PRL,Moessner1998PhysRevB.58.12049,Isakov2004PhysRevLett.93.167204,Henley05PRB,Henley2010ARCMP,Garanin1999PhysRevB.59.443,Rehn16PRL,Rehn17PRL,Benton16NComms,Taillefumier2017PhysRevX.7.041057,Yan20PRL,Benton21PRL,davier2023combined,Gembe2023arxiv}.
Even though CSLs tend to be unstable to perturbations at $T=0$, their large entropy {at low energies} can allow them to dominate the surrounding phase diagram at finite $T$. 
This makes them extremely relevant to the finite temperature physics of real frustrated systems. 
In addition, they can often be usefully thought of as parent states, or intermediate temperature limits, of quantum spin liquids which arise when  quantum fluctuations introduce dynamics between the classical ground states 
\cite{RVB_Fazekas,RK_1988,MS_RVB,Hermele2004PhysRevB.69.064404,Gingras2014RPPh,Sibille2018NatPh,Gaudet2019PhysRevLett.122.187201,Gao2019,Sibille2020,poree2023fractional}.

It is therefore important to understand and classify the spin liquids.
While classification schemes for QSLs have been successful, notably using the projective symmetry group \cite{Wen2002PhysRevB.65.165113}, and the modern perspective of gapped QSLs corresponding to a topological quantum field theory has led to further efforts in classifying QSLs as  symmetry-enriched topological  phases 
\cite{essin-hermele2013,Barkeshli2019PRB},
no similarly comprehensive classification exists for the CSLs.
Previous works have classified frustrated classical 
spin systems using constraint counting \cite{Moessner1998PhysRevB.58.12049}, linearization  around given spin configurations \cite{Roychowdhury2018PRB}, supersymmetric connections between models \cite{Roychowdhury-arXiv}
or using topological invariants built for 
specific cases \cite{Benton21PRL}. 
A scheme which generalizes across different models and
types of CSL, and which depends on the physics
of the CSL as a whole rather than individual spin configurations within it remains to be found.
Concretely, the question  whether CSLs have hidden topological properties like QSLs is an open one. Further, it is not clear how to place spin liquids with  algebraically-decaying and  exponentially-decaying correlations in a single scheme.

\begin{figure}[t!]
 \centering
 \includegraphics[width=0.95\columnwidth]{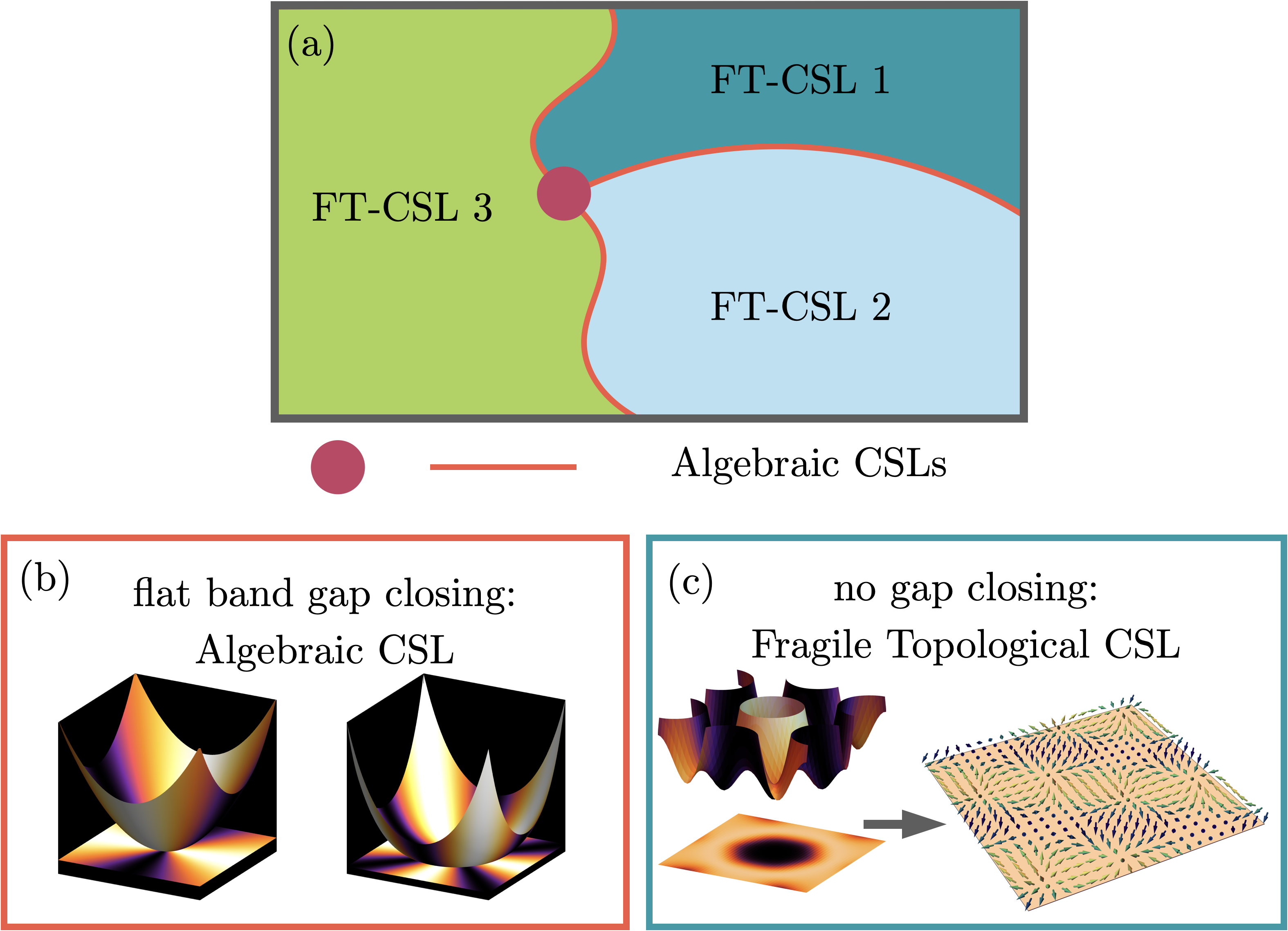} 
 \caption{(a) The landscape of CSLs consists of  Fragile Topological CSLs (FT-CSLs) whose boundaries are    Algebraic  CSLs.
 (b) 
 Algebraic CSLs feature gap closings between the bottom flat bands and higher dispersive bands of the spectrum of the exchange Hamiltonian. The band touching point determines the emergent Gauss's law.
 (c) 
 FT-CSLs have no such gap-closing points, being classified by their eigenvector homopoty.
\vspace{-2mm} }
 \label{Fig_overview}
\end{figure}

We present a classification scheme to address these issues, providing a concrete and relatively simple framework to distinguish different CSL states. Our work also offers practical tools  to diagnose known CSLs, as well as to construct new ones with desired properties.
We demonstrate in particular that certain CSLs can be characterised by a topological invariant that persists as long as the lowest flat bands of the spectrum of the Hamiltonian are separated by a gap from the higher dispersive band(s). 
We term these  ``fragile topological" classical spin liquids (FT-CSLs), as adding additional spins to the unit cell can render them topologically trivial without closing the spectral gap.

Algebraic spin liquids can be viewed as inhabiting the boundaries between the FT-CSLs where the spectral gap closes, illustrated schematically in Fig.~\ref{Fig_overview}(a).
This work presents a largely concept-based, non-technical account of the central narrative underpinning the classification scheme. 
We will present an extended discussion, including a detailed description alongside the more technical aspects, together with a broader set of new CSLs obtained within this scheme, in a long companion paper~\cite{Han2023arXivCLS2}.

\vspace{1mm}
\noindent\textbf{\textit{Classification.}}
Amongst classical spin models with continuous spins there exist a number of well-established CSLs
\cite{Villain1979,MC_pyro_PRL,Moessner1998PhysRevB.58.12049,Isakov2004PhysRevLett.93.167204,Garanin1999PhysRevB.59.443,Henley05PRB,Henley2010ARCMP,Rehn16PRL,Rehn17PRL,Benton16NComms,Taillefumier2017PhysRevX.7.041057,Yan20PRL,Benton21PRL,davier2023combined,Gembe2023arxiv}. 
Historically the first was the Heisenberg antiferromagnet on the pyrochlore lattice \cite{MC_pyro_PRL,Moessner1998PhysRevB.58.12049}, which exhibits an emergent U(1) gauge field in the low-energy description of its so-called Coulomb phase \cite{Isakov2004PhysRevLett.93.167204,Henley05PRB,Henley2010ARCMP,Castelnovo2012ARCMP}. 

It turns out that an adequate description of such models can often be attained in a self-consistent Gaussian (also known as soft-spin, or  large-$\mathcal{N}$) approximation \cite{Garanin1999PhysRevB.59.443,Isakov2004PhysRevLett.93.167204}, which is 
inspired by the Luttinger--Tisza method \cite{Luttinger1946PhysRev.70.954,Luttinger1951PhysRev.81.1015,Schmidt_2022}. This amounts to 
abandoning the hard spin-length constraint $|\mathbf{S}_i| = 1$, and replacing it with an average constraint $\langle \mathbf{S}^2\rangle = 1$. 
This results in a solvable Hamiltonian bilinear in spin variables, which can be diagonalised  in momentum space, yielding a spectrum with a band structure on which  
our classification is based~\footnote{Note that the spectrum is the property of the classical Hamiltonian, and not of a (specific) ground state. This is distinct from, for instance, the large-$S$ approach to quantum spin models where one expands the fluctuations around an ordered classical state to obtain the spin-wave spectrum.}.

The extensive ground state degeneracy of the CSL is reflected in the existence of (at least one) flat band at the bottom of the spectrum in momentum space -- because all the plane-wave states with a given momentum $\mathbf{q}$ in this band are  degenerate. The central concept behind our classification scheme is that the unique physics of different CSLs is encoded in their corresponding  band structures.
The most basic distinction 
is whether or not the bottom flat bands   are separated by a gap from the higher energy ones, illustrated in Fig.~\ref{Fig_overview}. Spectra with a band gap have short-range classical spin correlations. The absence of the gap, on the other hand, results in an algebraic CSL, of which the aforementioned U(1) Coulomb phase is one example. We shall formulate the classification of both these types of CSLs, starting with the latter.

\vspace{1mm}
\noindent
\textbf{\textit{Algebraic CSLs.}} 
We begin by examining algebraic CSLs. 
Due to the presence of gap-closing points, these systems exhibit spin correlations that decay algebraically. 
The well-known examples of   
Heisenberg antiferromagnet on the pyrochlore lattice \cite{AndersonPhysRev.102.1008,MC_pyro_PRL,Moessner1998PhysRevB.58.12049,Harris1997PhysRevLett.79.2554,Isakov2004PhysRevLett.93.167204,Isakov2005PhysRevLett.95.217201,Hermele2004PhysRevB.69.064404,Henley05PRB,Fennell2009Science,Henley2010ARCMP,Castelnovo2012ARCMP}, and also various other models \cite{Garanin1999PhysRevB.59.443,Rehn16PRL,Benton16NComms,Taillefumier2017PhysRevX.7.041057,Yan20PRL,Benton21PRL,davier2023combined,Gembe2023arxiv}, belong to this category.
While the gap-closing manifold can be more general, such as lines or surfaces \cite{Benton16NComms},
here we limit our discussion to gap-closing points only. 
By examining the eigenvector configuration around these points, we demonstrate how to extract a generalized Gauss's law, which plays a crucial role in describing the low-energy effective field theory of the CSL. The key steps of this procedure are outlined next.

We consider a system with $N$ spins per unit cell. The Hamiltonian for CSL systems can generally be expressed as a  \textit{constrainer Hamiltonian}:
\[ 
\mathcal{H} =     \sum_{\mathbf{R} \in \text{u.c.}}   [\mathcal{C} (\mathbf{R} )]^2 .
\] 
$\mathcal{C}(\mathbf{R})$, which we call a constrainer, involves  the sum of spins {with different coefficients in a local region}  around the unit cell located at $\mathbf{R}$ (see Eq.~\eqref{EQN_P_KGM} for an example). The Hamiltonian is the translationally invariant sum of such squared constrainers. 
The ground states of the system are spin configurations such that all constrainers are zero.
The
constrainer formalism is quite a general form of writing   Heisenberg Hamiltonians that realize CSLs.
Unless stated otherwise, the following discussion focuses on two spatial dimensions.

For simplicity, we assume there to be one constrainer per unit cell,
but its generalization is not difficult.
The spectrum of the Hamiltonian then contains $N-1$ bottom flat bands and one top dispersive band with a gap-closing point which we set to be at $\mathbf{q} = \mathbf{q}_0$.
The eigenvector $\mathbf{T}(\mathbf{q})$ of the top band,  computed as the Fourier transform of the constrainer (see example later), enables us to write down the $N\times N$ bilinear interaction matrix in momentum space 
\[
J_{ab}(\mathbf{q}) = T_a T_b^*(\mathbf{q}),\quad a ,b = 1,\dots, N,
\] 
explicitly featuring $N-1$ bottom flat bands at $\omega = 0$ and a top band with a dispersion relation $\omega_T(\mathbf{q}) = |\mathbf{T}(\mathbf{q})|^2$.

Let us analyze the behavior of this eigenvector for small wave-vectors $ \mathbf{k} = \mathbf{q}- \mathbf{q}_0$.
The gap-closing condition implies that $\mathbf{T}(\mathbf{q}_0) =\mathbf{0}$.
Hence we can express its components,  denoted as ${T}_a(\mathbf{q}_0+\mathbf{k})$, as  Taylor expansions in $k_x$ and $k_y$ without the zeroth order constants. The leading  terms in this polynomial expansion are
\[ 
\label{EQN_T_expansion_general}
T_a (\mathbf{q}_0 + \mathbf{k} )  
 = \sum_{j=0}^{m_a} c_{aj} (-ik_x)^{j} (-ik_y)^{m_a-j}, 
\]
where $m_a$ is the degree of the leading order terms in  ${T}_a(\mathbf{q})$ and the constants $c_{aj}$ are given by the Taylor expansion. Ground states are associated with the ($N\!-\!1$)-dimensional space of (complex-valued) eigenvectors ${\mathbf{S}}$ orthogonal to $\mathbf{T}(\mathbf{q}_0 + \mathbf{k})$, which satisfy
\[
\label{EQN_Gauss_general}
\mathbf{T}^\ast(\mathbf{q}_0+ \mathbf{k}) \cdot {\mathbf{S}}=  \sum_{a=1}^N \sum_{j=0}^{m_a} c_{aj}^\ast (ik_x)^{j} (ik_y)^{m_a-j}{S}_a = 0.   
\]
Performing the inverse Fourier transformation into real space yields the generalized Gauss's law:
\[
\label{EQN_Gauss_general_real_3}
\rho = \sum_{a=1}^N\left(\sum_{j=0}^{m_a} c_{aj}^\ast (\partial_x)^{j} (\partial_y)^{m_a-j} S_a \right) \equiv \sum_{a=1}^ND_a^{(m_a)} S_a ,
\]
and the ground state condition is $\rho =0$.

Crucially, this analysis also yields 
the equal-time spin structure factor, 
the intensity distribution of which,
$1 - |\sum_a T_a|^2/ \mathbf{T} ^2$,
exhibits singular patterns at $\mathbf{q_0}$, known as pinch points~\cite{Henley05PRB,Isakov2004PhysRevLett.93.167204,Prem18PRB,Hart2022PhysRevB,PhysRevB.98.140402}[Fig.~\ref{Fig_overview}(b)].

Such Gauss's laws play a central role in describing the properties of the ground state manifold of algebraic CSLs. Specifically, the long-wavelength expansion results in an effective Hamiltonian given by $\mathcal{H}_\text{eff} =  (\sum_{a=1}^ND_a^{(m_a)} S_a)^2$, which properly captures the algebraic spin correlation of the system.  
The generalized Gauss's laws can also give rise to non-trivial physics, such as multipole conservations and fracton charges, which have garnered attention in various fields of physics
\cite{Xu2006PhysRevB,PretkoPRB16,PretkoPRB17,NandkishoreReview,Pretko-review2020}.
Equipping this Hamiltonian with quantum dynamics provides the starting point for building the emergent (generalized) electrodynamics describing the corresponding QSL.
In addition, the presence of multiple gap-closing points allows for the coexistence of different generalized Gauss's laws that describe the same ground state manifold, depending on the background wave-vectors $\mathbf{q}$ in the long-wavelength limit.

{We can now distinguish different algebraic CSLs by comparing their gap-closing points.
Concretely, two algebraic CSLs can be viewed as in the same class if one can adiabatically deform the constrainer Hamiltonian and turn the Gauss's law of one CSL into that of the other, without going through singular processes of merging/splitting/lifting some of these points. 
On the other hand, two algebraic CSLs are categorically different, if they have a different number of gap-closing points, or the associated Gauss's laws involve a different number of effective electric field degrees of freedom, or a different order of $\partial_x, \partial_y$.
These gap-closing points cannot be made identical without going through certain singular transitions.}
Below, we provide an example to demonstrate how the Gauss's law is extracted from a concrete model, and how the merging/splitting of the gap-closing point happens.

\vspace{1mm}
\noindent\textbf{\textit{Fragile Topological CSLs.}} 
We now turn to the second category of CSLs, FT-CSLs. They are characterized by bottom flat bands that are completely separated from other bands above them by a gap, resulting in exponentially decaying spin correlations \cite{Rehn17PRL}.

The FT-CSLs  can be further classified based on the homotopy class of the bottom band eigenvector configuration,  which can only change when the system undergoes a gap-closing topological phase transition.
{Hence the phase boundaries of such FT-CSLs  are inhabited by the algebraic CSLs. This is similar to the concept of topological band transitions, where the Chern number of a band cannot change without closing a gap.}

The specific classification scheme for 2D   FT-CSLs with $N$ sub-lattice sites and one  (or, analogously, $N-1$, see below) bottom flat bands works as follows. 
Consider a normalized single bottom flat band eigenvector, denoted by $\hat{\mathbf{B}}(\mathbf{q})$ (or the top band eigenvector $\hat{\mathbf{T}}(\mathbf{q})$ if there is only one top band). The components of $\hat{\mathbf{B}}(\mathbf{q})$ are generally complex, but for certain symmetries they can be also real. 
The flatness of the   band ensures that it has a zero Chern number \cite{Chen2014}, which means the eigenvector $\hat{\mathbf{B}}(\mathbf{q})$ is smoothly defined over the entire Brillouin zone (BZ). 
Since the BZ is a 2-torus, $\hat{\mathbf{B}}(\mathbf{q})$ defines a map from the torus to the target space of $\CP^{N-1}$ or $\RP^{N-1}$:
\[
\hat{\mathbf{B}}(\mathbf{q}): {T}^2 \rightarrow \CP^{N-1} (\text{or }\RP^{N-1});\  \mathbf{q} \mapsto \hat{\mathbf{B}}(\mathbf{q}).
\]

This eigenvector can still ``wind" nontrivially over the BZ -- captured by the homotopy class of the corresponding map  $[{T}^2, \CP^{N-1} ]$ or $[{T}^2, \RP^{N-1} ]$. The homotopy class can only change at gap-closing points, where $\hat{\mathbf{B}}(\mathbf{q})$ becomes ill-defined. 

In the case where $\hat{\mathbf{B}}(\mathbf{q})$ is complex, we can use the fact that the complex projective space is simply connected ($\pi_1(\CP^{N-1})=0$) for any $N-1\geq 1$ to obtain:
\[
\label{eqn:homoT2RPn}
[{T}^2, \CP^{N-1} ] \cong \pi_2 (\CP^{N-1} ) = \mathbb{Z}.
\] 

For real-valued $\hat{\mathbf{B}}(\mathbf{q})$, the homotopy class of $[{T}^2, \RP^{N-1} ]$ does not have a simple formula. However, in a special case when one can consistently assign directions to the $\RP^{N-1}$ eigenvectors over the boundary condition of the BZ, the homotopy group simplifies:
\[
\label{eqn:homoT2Sn}
\left. [{T}^2, S^{N-1} ]\right|_{b.c.} \cong 
\pi_2 (S^{N-1} ) = \begin{cases}
 \mathbb{Z} & \text{if $N-1 = 2$}\\
 0 & \text{if $N-1 \ge 3$}
 \end{cases} .
\]
The only non-trivial case is when $N=3$, which is the skyrmion number on the torus, $Q_\mathsf{sk}$, given by
\[\label{EQN_skyrmion}
Q_\mathsf{sk}=\frac{1}{4 \pi} \int_{\rm BZ} \mathrm{d}^2{\bf q}\  \hat{\mathbf{B}}({\bf q}) \cdot
\left(
\frac{\partial \hat{\mathbf{B}}({\bf q})}{\partial q_x} 
\times 
\frac{\partial \hat{\mathbf{B}}({\bf q})}{\partial q_y}
\right) .
\]
In the case of three-dimension models, we need to compute $[{T}^3, \CP^{N-1} ]$ or $[{T}^3, \RP^{N-1} ]$ instead.

Note that the homotopy class is a \textit{fragile} topological quantity in the following sense: if new spins are added to each unit cell and interact with the original spins, the  non-trivial homotopy class may become trivial in the new model. 
By padding each unit cell with  auxiliary spins, adiabatically tuning the CSL Hamiltonian and then decoupling them, one can change the homotopy class without closing the spectral gap.

\begin{figure}[t!]
 \centering
 \includegraphics[width=0.82\columnwidth]{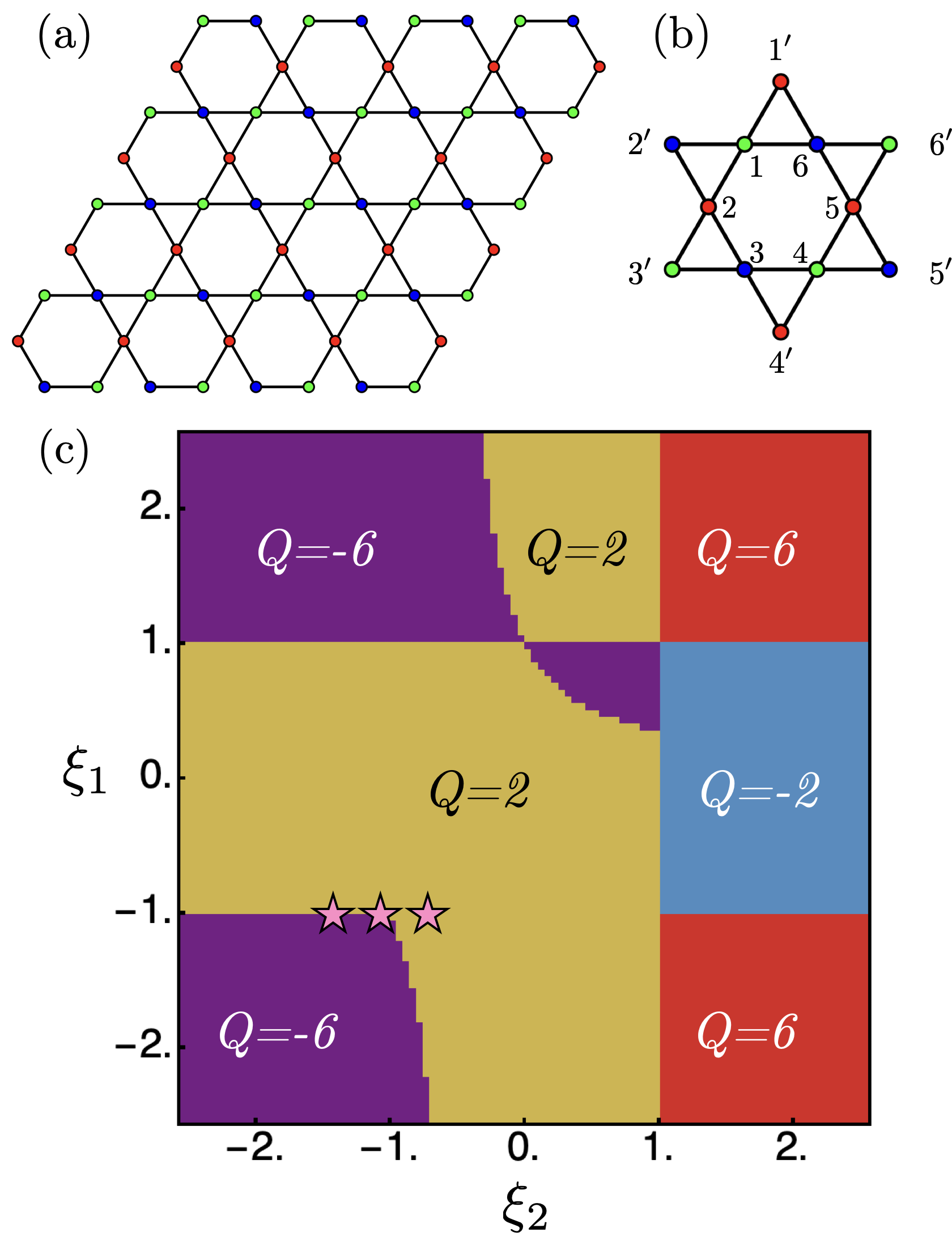} 
 \caption{
 The kagome model of $\mathcal{H}_\text{KGM}$ in Eq.~\eqref{EQN_H_KGM}
 (a) The Kagome lattice with different sublattice sites colored differently.
 (b) The  sites for the constrainer $\mathcal{C}_\text{KGM}$ defined in Eq.~\eqref{EQN_P_KGM}. 
 (c) The phase diagram of the model Eq.~\eqref{EQN_H_KGM}.
 Different  colored regions are phases of fragile topological CSL with different skyrmion numbers $Q$. 
 The three pink stars are the parameter sets shown and Fig.~\ref{Fig_dispersion}.
\vspace{-4mm} }
 \label{Fig_winding_phase}
\end{figure}

\vspace{1mm}
\noindent\textbf{\textit{Example.}} 
We illustrate the classification scheme with the concrete example of  a kagome lattice [Fig.~\ref{Fig_winding_phase}(a)] model. The constrainer Hamiltonian reads:
\begin{eqnarray} \label{EQN_H_KGM}
\mathcal{H}_\text{KGM} &=&     \sum_{\mathbf{R} \in \text{u.c.}}   [\mathcal{C}_\text{KGM} (\mathbf{R} )]^2 
\\
\label{EQN_P_KGM}
\mathcal{C}_\text{KGM} (\mathbf{R} )
&=& \sum_{i  = 1}^{6} S_i
+ \xi_1 \sum_{j  =2'3'5'6'}  S_{j} + \xi_2 \sum_{j  =  1' 4'}  S_{j}.
\end{eqnarray}
Here, the sites $1,\dots ,6$ and $1',\dots ,6'$ in $\mathcal{C}_\text{KGM} (\mathbf{R})$ are labeled in Fig.~\ref{Fig_winding_phase}(b) for the hexagonal star located at $\mathbf{R}$.
The case of $\xi_1 = \xi_2 = 0$ was studied in Ref.~\cite{Rehn17PRL}.

Our model has one constrainer per unit cell and three sub-lattice sites, leading to a spectrum that consists of two degenerate bottom flat bands and a top dispersive band. The eigenvector of the top band can be expressed as the    Fourier transform  of $\mathcal{C}_\text{KGM} (\mathbf{R})$:
\[ \label{EQN_Top_band_KGM}
{\bf T} ({\bf q})
=
\begin{pmatrix}
\cos(\sqrt{3} q_x) + \xi_2 \cos(3 q_y)\\
\cos\left( -\frac{\sqrt{3}}{2} q_x + \frac{3}{2} q_y \right) + \xi_1 \cos(- 3\frac{\sqrt{3}}{2} q_x - \frac{3}{2} q_y) \\
\cos\left( -\frac{\sqrt{3}}{2} q_x - \frac{3}{2} q_y \right) + \xi_1 \cos(3\frac{\sqrt{3}}{2} q_x - \frac{3}{2} q_y) 
\end{pmatrix},
\]
and its dispersion is  $\omega(\mathbf{q} ) = |{\bf T}  ({\bf q})|^2$.

\begin{figure}[t!]
 \centering
 \includegraphics[width=0.84\columnwidth]{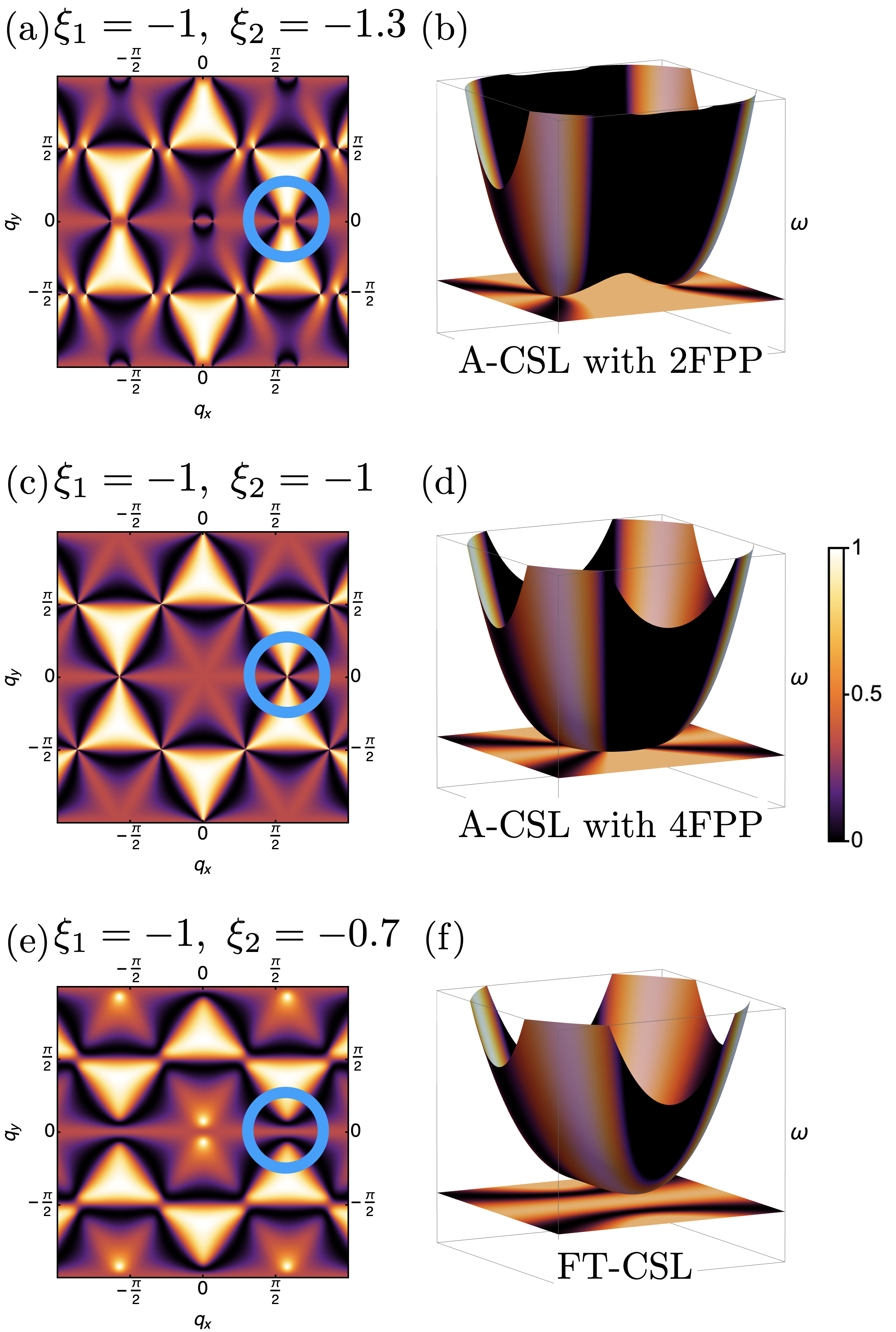} 
 \caption{Spin structure factor and spectrum of three parameter sets (pink stars in Fig.~\ref{Fig_winding_phase}) of the model Eq.~\eqref{EQN_H_KGM}, highlighting the merging and lifting of the gap-closing point (blue circle), which indicates transition between algebraic  CSLs  and fragile topological CSLs.
 (a) Spin Structure for parameter $\xi_1 = -1, \xi_2 = -1.3$.
 (b)
 The two gap-closing points in its spectrum each hosts a 2-fold pinch point (2FPP).
 (c) Spin Structure   for parameter $\xi_1 = -1, \xi_2 = -1$.
 (d) The previous two gap-closing points merge and form a single gap-closing point with 4FPP.
 (e) Spin Structure   for parameter $\xi_1 = -1, \xi_2 = -0.7$.
 (f) The     gap-closing point lifts up and opens up the gap.
\vspace{-4mm} }
 \label{Fig_dispersion}
\end{figure}

We compute $Q_\mathsf{sk}$, Eq.\eqref{EQN_skyrmion}, to determine the homotopy class of the eigenvector $\mathbf{T}(\mathbf{q})$ of the top dispersive band (instead of the bottom band eigenvector $\mathbf{B}(\mathbf{q})$, which amounts to replacing $\mathbf{B} \to \mathbf{T}$ in the above formalism). Tuning the two parameters $\xi_{1,2}$ yields the diverse phases shown in Fig.~\ref{Fig_winding_phase}(c), labeled by their skyrmion numbers. 

The boundaries of these topological phases, corresponding to the gap closing between the bands, host various algebraic CSLs. Their emergent Gauss's laws are obtained by substituting the top band eigenvector in Eq.~\eqref{EQN_Top_band_KGM} into Eqs.~(\ref{EQN_T_expansion_general},\ref{EQN_Gauss_general_real_3}).
For $\xi_1 = \xi_2 = -1$, the gapless point $\mathbf{k} = (0,\pi/\sqrt{3}) + (k_x, k_y)$  exhibits the Gauss's law 
\[
\rho = 2\sqrt{3}\partial_x \partial_y (S_1 - S_3)
+   \partial_x^2 (2S_1 + S_2  + 2S_3) 
- 3 \partial_y^2 S_2  .
\]
This can be recast  as a symmetric rank-2 U(1) Gauss's law of a scalar charge \cite{Xu2006PhysRevB,PretkoPRB16,PretkoPRB17}\[
\partial_\alpha \partial_\beta  E_{\alpha \beta} = 0,
\]
where the electric field is a rank-2 symmetric tensor
\[
E_{\alpha \beta}=
\begin{pmatrix}
 2S_1 +  S_2  + 2S_3 &  \sqrt{3}(S_1 - S_3) \\
 \sqrt{3}(S_1 - S_3) &
  -3 S_2
\end{pmatrix}.
\]
In the spin structure factor, this is characterized by  the 4-fold pinch point (4FPP) \cite{Prem18PRB} [Fig.~\ref{Fig_dispersion}(c,d)].

Decreasing the value of $\xi_2$ moves along the phase boundary, but the spectrum of the Hamiltonian and the emergent Gauss's law changes, with the 4-fold pinch point (4FPP) splitting into two 2-fold pinch points  [Fig.~\ref{Fig_dispersion}(a,b)]: this  transition between algebraic CSLs involves the merging and splitting of gap-closing points.

By contrast, upon increasing $\xi_2>-1$,
the gap-closing point is lifted, and a gap opens up between the flat  and dispersive bands. This yields a FT-CSL, as shown in Fig.~\ref{Fig_dispersion}(e,f).
Other phase boundaries of algebraic CSLs  are also interesting, but we will not delve into them extensively. We do note that with our classification methodology, analysing these phase boundaries is now a straightforward, basic algebraic calculation. 

\vspace{1mm}
\noindent\textbf{\textit{Summary.}}
In this work, we have presented a classification scheme for classical spin liquids, which we divide into 
 two broad categories: algebraic and \ft CSLs. 
In our extended companion paper \cite{Han2023arXivCLS2}, we present a  comprehensive analysis of the classification scheme, including aspects omitted here such as higher-dimensional CSLs, more complex band-closing structures, and connection to flat band theories. Additionally, we construct a variety of new models using the constrainer Hamiltonian formalism to illustrate the different aspects of the classification scheme, and the new physics arising from it.

\vspace{2mm}
\textit{Acknowledgements.} H.Y. and A.H.N. were supported by the U.S. National Science Foundation Division of Materials Research under the Award DMR-1917511.
This work was in part
supported by the Deutsche Forschungsgemeinschaft under
grant SFB 1143 (project-id 247310070) and the cluster of excellence 
ct.qmat (EXC 2147, project-id 390858490).

\vspace{2mm}
\textit{Note.} During completion of this manuscript, a
preprint by Davier {\it et al.} \cite{davier2023combined} appeared, which independently presents results regarding the classification of CSLs.

\bibliography{reference}
  
\end{document}